\documentclass[11pt,a4paper]{article}

\usepackage[utf8]{inputenc}
\usepackage[T1]{fontenc}
\usepackage{amsmath,amssymb}
\usepackage{bm}
\usepackage{graphicx}
\usepackage{booktabs}
\usepackage{hyperref}
\usepackage{url}
\usepackage[numbers,sort&compress]{natbib}
\usepackage{authblk}
\usepackage[margin=2.5cm]{geometry}
\usepackage{caption}
\usepackage{float}
\usepackage{enumitem}
\usepackage{xcolor}

\hypersetup{
  colorlinks=true,
  linkcolor=blue,
  citecolor=blue,
  urlcolor=blue
}

\newcommand{\Us}{U^{*}}
\newcommand{\Uref}{U^{*}_{\mathrm{ref}}}
\newcommand{\dU}{\Delta U^{*}}
\newcommand{\vect}[1]{\bm{#1}}
\newcommand{\doi}[1]{\href{https://doi.org/#1}{doi:#1}}
\newcommand{\orcid}[1]{\textsuperscript{\href{https://orcid.org/#1}{\scriptsize ORCID}}}

\title{\textbf{Transducer Placement and the Limits of a Four-State Reduced Model
in Post-Flutter Piezoelectric Energy Harvesting from a Pitch-Plunge-Flap
Aerofoil}}

\author[1]{Nikolaos~D.~Tantaroudas\orcid{0000-0001-6727-1014}\thanks{Corresponding
author: \texttt{nikolaos.tantaroudas@iccs.gr}. ORCID iDs:
N.~D.~Tantaroudas 0000-0001-6727-1014, I.~Karachalios 0009-0000-8581-8335.}}
\author[2]{Ilias~Karachalios\orcid{0009-0000-8581-8335}}
\author[3]{Andrew~J.~McCracken}

\affil[1]{Institute of Communications and Computer Systems (ICCS), National
Technical University of Athens, Iroon Polytechniou 9, 15773 Zografou, Greece}
\affil[2]{Department of Environmental Sciences, University of Thessaly, Gaiopolis
Campus, Larissa-Trikala Ring Road, 41500 Larissa, Greece}
\affil[3]{DASKALOS APPS, 183 Rue de l'Abb\'{e} Griffon, 01960 P\'{e}ronnas,
France}

\date{}

\begin{document}

\maketitle

\begin{abstract}
Aeroelastic flutter is normally a failure mode to be designed against, yet the
limit-cycle oscillations (LCOs) that follow it convert flow energy into sustained
structural motion that a piezoelectric transducer can turn into electrical power.
A transducer is embedded in a three-degree-of-freedom pitch-plunge aerofoil with a
finite-mass trailing-edge flap and unsteady strip-theory aerodynamics, giving a
fifteen-state electro-aeroelastic system with a cubic hardening pitch spring, and
the system is reduced by biorthonormal projection of the Taylor-expanded residual
onto eigenvectors of the coupled Jacobian. The testbed is deliberately low order,
so that every reduced prediction can be checked against the full-order system it
replaces. The degree of freedom that carries the transducer is a first-order
design variable, since it sets both the sign of the shift in the flutter boundary
and the magnitude of the harvested power. The error of the reduced model is
dominated not by the size of the retained basis but by how the reduced operator is
made to depend on flow speed. Expanding the retained eigenvalues leaves the reduced
operator block-diagonal, with no coupling between the retained modes at any order
of the expansion, whereas re-projecting the exact Jacobian onto the same frozen
basis supplies that coupling and restores the structural response without
enlarging the basis. The error that remains lies in the harvested voltage, which the
equilibrium eigenvectors carry in a fixed proportion to the motion, and it is not
removed by adding the electrical modes to the basis, whereas the flap
deflection, which the same basis also misses, is recovered by retaining the flap
modes. The accuracy of a harvester reduced-order model must therefore be reported
on the harvested quantity itself. 
\end{abstract}


\section{Introduction}
\label{sec:intro}

Aeroelastic instability is conventionally a constraint, in that the flutter
boundary is a speed to be designed away from and the LCOs that follow it are a
fatigue and handling problem. The same instability is, however, a persistent
conversion of flow energy into structural motion, part of which a piezoelectric
transducer can convert into electrical power. Erturk and co-workers generated
$10.7$~mW at the linear flutter speed of a piezoaeroelastic airfoil and noted that
the transduction itself displaces the flutter boundary~\citep{erturk2010}.
Because a linear system at its flutter boundary has no bounded amplitude,
practical devices operate post-critically, on an LCO set by a structural or
aerodynamic nonlinearity. Power has been extracted from the LCOs of a flexible
beam in axial flow~\citep{dunnmon2011}, and combining free play with cubic
hardening widens the range of flow speeds over which a bounded amplitude
persists~\citep{sousa2011}. The piezoaeroelastic harvester has been
modelled~\citep{abdelkefi2011}, piezoelectric and electromagnetic transduction
have been compared on airfoil-based devices~\citep{demarqui2013}, and a
flutter-driven harvester has been built and tested~\citep{bryant2011}. The source
considered here is a deterministic limit cycle rather than broadband random
excitation~\citep{adhikari2009}. The field is reviewed in~\citep{abdelkefi2016}.

Designing such a harvester is a parameter-sweep problem. The harvested power
depends on the flow speed, the electrical load, the electromechanical coupling and
the degree of freedom that carries the transducer, and it is defined only on a
converged limit cycle, so every point of the sweep needs a nonlinear time march.
For high-fidelity aeroelastic models that sweep is prohibitive, which motivates a
reduced-order model. The difficulty specific to harvesting is that a basis chosen
to capture the dominant structural mode need not capture the electrical state that
carries the harvested power.

The present work builds on a line of nonlinear reduced-order modelling developed by
the authors. The Taylor-expansion residual formulation with biorthonormal
eigenvector projection and matrix-free higher-order operators was introduced for
gust loads analysis~\citep{daronch2013gust} and control
applications~\citep{daronch2013control}, and extended to adaptive aeroelastic
control~\citep{tantaroudas2014adaptive}, to flutter suppression validated from
simulation to wind-tunnel testing~\citep{daronch2014flutter,papatheou2013}, and
to very flexible aircraft~\citep{tantaroudas2015vfa,tantaroudas2017vfa}. The
pitch-plunge aerofoil with a finite-mass trailing-edge flap used here is the
testbed characterised experimentally and numerically in~\citep{fichera2014}. More
recent work has applied the same reduction to coupled aeroelastic-flight-dynamic
systems~\citep{tantaroudas2026coupled}, worst-case gust
identification~\citep{tantaroudas2026gust} and $H_\infty$ gust load
alleviation~\citep{tantaroudas2026hinf}, and the influence of the aerodynamic
modelling level is examined in~\citep{daronch2014aero}.

The contribution is threefold. First, the formulation
of~\citep{daronch2013gust,daronch2013control} is extended from load alleviation
and flutter suppression, where aeroelastic energy is dissipated, to harvesting,
where it is extracted, and the mounting degree of freedom is shown to change both
the sign of the flutter-boundary shift and the harvested power by a factor of
four and a half. Second, the error of the reduced model is decomposed. It is
dominated not by the size of the retained basis but by how the reduced operator is
made to depend on flow speed, and separating the two reduces the peak
pitch-amplitude error from $46\%$ to $3.4\%$ without enlarging the basis. Third,
the error that remains is shown to lie in the harvested quantity rather than in
the pitch and plunge response, and to be largely insensitive to adding the
electrical modes to the basis, which moves the voltage error only from $-22\%$ to
$-16\%$. The testbed is deliberately low order, so that every reduced prediction
can be checked against the full-order system it replaces. \textbf{NFOM} and
\textbf{NROM} denote the nonlinear full- and reduced-order models, while \textbf{FOM} and \textbf{ROM} are reserved for their linearisations.

\section{Electro-aeroelastic model}
\label{sec:model}

\subsection{Structure and unsteady aerodynamics}

\begin{figure}[H]
\centering
\includegraphics[width=0.86\textwidth]{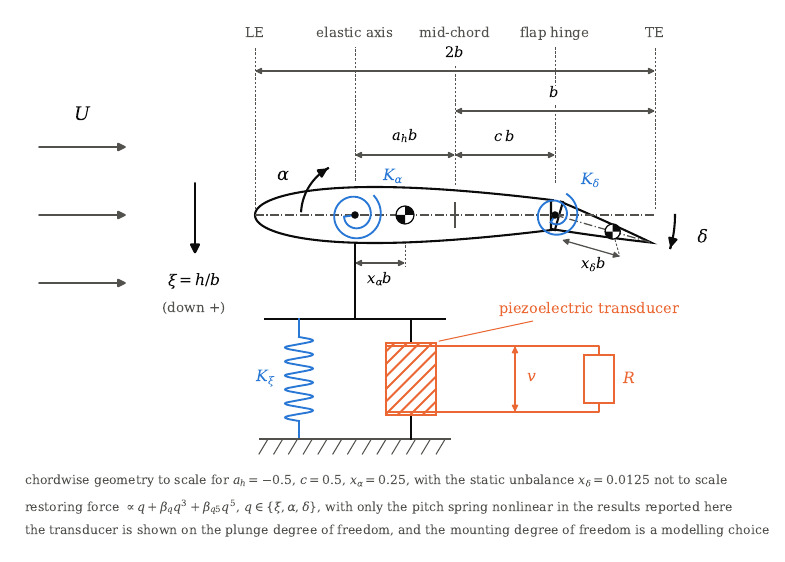}
\caption{The three-degree-of-freedom section with its plunge-mounted piezoelectric
transducer and resistive shunt. Stations are $x/b$ from the midchord.}
\label{fig:schematic}
\end{figure}

The testbed is a two-dimensional aerofoil section free to plunge, to pitch about an
elastic axis and to rotate a finite-mass trailing-edge flap about its hinge
(Figure~\ref{fig:schematic}). Lengths are scaled by the semi-chord $b$, time as
$\tau = Ut/b$, and flow speed as the reduced velocity $\Us = U/(b\omega_\alpha)$.
The generalised coordinates are $\vect{q} = (\xi,\alpha,\delta)^{\mathsf T}$, with
$\xi = h/b$ the plunge, positive downward, $\alpha$ the pitch, positive nose-up,
and $\delta$ the flap deflection, positive trailing-edge-down, and primes denote
$\mathrm{d}/\mathrm{d}\tau$. With $a_h = -0.5$ and $c = 0.5$ the elastic axis lies
at the quarter chord and the hinge at $75\%$ chord. The elastic axis then
coincides with the thin-aerofoil aerodynamic centre, so the circulatory lift exerts
no moment about it. The structural stiffness is diagonal and the structural damping
is zero. The equations of motion are
\begin{equation}
\vect{M}\vect{q}'' + \vect{C}\vect{q}' + \vect{K}\vect{q}
+ \vect{f}_{\mathrm{nl}}(\vect{q}) + \vect{W}\vect{w}
+ \kappa\,\vect{e}_{j}\,v = \vect{0}.
\label{eq:struct}
\end{equation}
The $3\times3$ matrices $\vect{M}$, $\vect{C}$ and $\vect{K}$ carry the structural
inertia, damping and stiffness together with the non-circulatory and quasi-steady
circulatory aerodynamic terms. Their entries are assembled from the mass ratio
$\mu$, the frequency ratios, $a_h$, $c$, the radii of gyration, the static
unbalances and the Theodorsen constants $T_1$ to $T_{13}$, following the
pitch-plunge-flap formulation of Lee and co-workers~\citep{lee1999} as implemented
in~\citep{tantaroudas2015phd}. The vector $\vect{w}$ holds the eight aerodynamic
lag states of Eqs.~\eqref{eq:wagner} and~\eqref{eq:kussner}, $\vect{W}$ maps them
into the equations of motion, and the last term is the electromechanical
back-coupling of the transducer, with $\vect{e}_j$ selecting the degree of freedom
that carries it. The restoring forces are polynomial,
\begin{equation}
\vect{f}_{\mathrm{nl}}(\vect{q}) =
\begin{bmatrix}
(\omega/\Us)^{2}\left(\beta_{\xi}\xi^{3}+\beta_{\xi5}\xi^{5}\right) \\[2pt]
(1/\Us)^{2}\left(\beta_{\alpha}\alpha^{3}+\beta_{\alpha5}\alpha^{5}\right) \\[2pt]
(\omega_{\delta}/\Us)^{2}\left(\beta_{\delta}\delta^{3}+\beta_{\delta5}\delta^{5}\right)
\end{bmatrix},
\label{eq:nonlin}
\end{equation}
with $\omega$ and $\omega_\delta$ the plunge/pitch and flap/pitch frequency
ratios. The quintic terms are zero throughout, and only the pitch spring is
nonlinear. Because the restoring forces carry no quadratic term, the quadratic
operator of the reduction in Section~\ref{sec:rom} vanishes identically.

The parameters used are given below, namely
$\mu = 100$, $\omega = 0.2$, $\omega_\delta = 3.5$, $a_h = -0.5$, $x_\alpha = 0.25$,
$r_\alpha = 0.5$, $c = 0.5$, $r_\delta = 0.0791$ and $x_\delta = 0.0125$. The cubic
pitch stiffness $\beta_\alpha = 100$ holds the limit cycle below five degrees of
pitch, inside the attached-flow validity of the aerodynamics.

The unsteady aerodynamics are strip theory in indicial form, with Wagner's and
K\"ussner's functions in the usual two-exponential fit,
\begin{equation*}
\phi(\tau) = 1-\psi_1 e^{-\varepsilon_1\tau}-\psi_2 e^{-\varepsilon_2\tau},
\qquad
\psi(\tau) = 1-\psi_3 e^{-\varepsilon_3\tau}-\psi_4 e^{-\varepsilon_4\tau},
\end{equation*}
with $\psi_1=0.165$, $\psi_2=0.335$, $\varepsilon_1=0.0455$,
$\varepsilon_2=0.300$ and $\psi_3=0.5792$, $\psi_4=0.4208$,
$\varepsilon_3=0.1393$, $\varepsilon_4=1.802$. Each generalised coordinate
contributes one lag state per Wagner exponent,
\begin{equation}
w_{i}' = q_{k(i)} - \varepsilon_{\sigma(i)}\,w_{i},
\qquad i = 1,\dots,6,
\label{eq:wagner}
\end{equation}
with $w_1,w_2$ driven by $\alpha$, $w_3,w_4$ by $\xi$ and $w_5,w_6$ by $\delta$,
each pair with rates $\varepsilon_1,\varepsilon_2$. Gust penetration adds two
states, one per K\"ussner exponent,
\begin{equation}
w_{7}' = -\varepsilon_{3}\,w_{7} + w_{g},
\qquad
w_{8}' = -\varepsilon_{4}\,w_{8} + w_{g},
\label{eq:kussner}
\end{equation}
with gust velocity $w_g = 0$ throughout. The gust states are kept so that the model
is the solver's own fifteen-state layout, and they are inert in every result
reported here. They matter to the reduction for one reason only, that one of their
poles lies close to the transducer pole, so the two eigenvectors mix
(Section~\ref{sec:rom}). The flap equation receives no gust contribution, a
simplification that is immaterial here because no gust is applied.

\subsection{Piezoelectric transduction}
\label{sec:piezo}

The transducer is a lumped element on one structural degree of freedom, with
linear electromechanical coupling $\theta$ and capacitance $C_p$, shunted by a
resistive load $R$, and in Figure~\ref{fig:schematic} it acts in parallel with the
plunge spring and has no stiffness of its own. Scaling its voltage as $v = C_pV/(\theta b)$, the charge balance
of the circuit becomes a first-order equation driven by the velocity of the
mounted degree of freedom,
\begin{equation}
v' = q_{j}' - \lambda v,
\qquad
\lambda = \frac{1}{\Us r},
\label{eq:circuit}
\end{equation}
and the harvester is characterised by exactly two dimensionless groups,
\begin{equation}
\chi = \frac{\theta^{2}}{k_{j} C_{p}},
\qquad
r = R\,C_{p}\,\omega_{\alpha},
\qquad
\kappa = \chi\,k_{j},
\label{eq:groups}
\end{equation}
where $k_j$ is the linear stiffness of the mounted degree of freedom, equal to
$(\omega/\Us)^2$ for plunge, $1/\Us{}^2$ for pitch and $(\omega_\delta/\Us)^2$ for
flap. The instantaneous power in the load, and its average over a whole number of
cycles, are given:
\begin{equation}
\bar{p} = \frac{\chi v^{2}}{r},
\qquad
\langle\bar{p}\rangle = \frac{1}{T}\int_{0}^{T}\frac{\chi v^{2}}{r}\,\mathrm{d}\tau.
\label{eq:power}
\end{equation}
Because Eq.~\eqref{eq:power} is normalised by $k_j$, powers from different
mountings are not the same physical quantity, so where mountings are compared they are
referred to the common reference $P/(mb^2\omega_\alpha^3)$ by multiplying by
$\omega^2$ for plunge, $r_\alpha^2$ for pitch or $r_\delta^2\omega_\delta^2$ for
flap. With the circuit open the voltage follows the displacement of the mounted
degree of freedom, so the reaction $\kappa v$ then acts as an added stiffness.
Equation~\eqref{eq:circuit} adds a single state, and collecting
$\vect{x} = (\vect{q},\vect{q}',w_1,\dots,w_8,v)$ gives the coupled system as a
fifteen-state autonomous problem,
\begin{equation}
\vect{x}' = \vect{F}(\vect{x};\Us) = \vect{A}(\Us)\vect{x} + \vect{f}(\vect{x};\Us),
\qquad \vect{x}\in\mathbb{R}^{15},
\label{eq:fom}
\end{equation}
whose linear part $\vect{A}(\Us)$ is the Jacobian at the origin.

\section{Nonlinear reduced-order model}
\label{sec:rom}

Following~\citep{daronch2013gust,daronch2013control}, the residual of
Eq.~\eqref{eq:fom} is expanded about the equilibrium,
\begin{equation}
\vect{F}(\vect{x}) = \vect{A}\vect{x}
+ \tfrac{1}{2}\vect{B}(\vect{x},\vect{x})
+ \tfrac{1}{6}\vect{C}(\vect{x},\vect{x},\vect{x})
+ \mathcal{O}(\|\vect{x}\|^{4}),
\label{eq:taylor}
\end{equation}
with $\vect{B}$ and $\vect{C}$ the second- and third-order Fr\'echet derivatives,
evaluated matrix-free by directional finite differences of $\vect{F}$ along the
retained basis vectors, so no high-order tensor is assembled in the full space.
Retaining $m$ complex conjugate eigenpairs of $\vect{A}(\Uref)$, with
biorthonormal right and left eigenvectors $\vect{\Phi}$ and $\vect{\Psi}$, and
writing $\vect{x} = 2\,\mathrm{Re}(\vect{\Phi}\vect{z})$ gives the real basis pair
\begin{equation*}
\vect{V} = [\,2\,\mathrm{Re}\,\vect{\Phi}\;\;{-}2\,\mathrm{Im}\,\vect{\Phi}\,],
\qquad
\vect{W} = [\,\mathrm{Re}\,\vect{\Psi}\;\;{-}\mathrm{Im}\,\vect{\Psi}\,],
\qquad
\vect{W}^{\mathsf T}\vect{V} = \vect{I},
\end{equation*}
and the reduced system in $2m$ real coordinates
\begin{equation}
\vect{z}' = \vect{A}_{r}(\Us)\,\vect{z}
+ \tfrac{1}{6}\,\vect{C}_{r}(\vect{z},\vect{z},\vect{z}),
\qquad
\vect{C}_{r}(\vect{y}_1,\vect{y}_2,\vect{y}_3)
= \vect{W}^{\mathsf T}\vect{C}(\vect{V}\vect{y}_1,\vect{V}\vect{y}_2,\vect{V}\vect{y}_3).
\label{eq:reduced}
\end{equation}

Two pairs are retained, $m = 2$, giving four real states. The first is
the least stable pair, the flutter mode, a coalescence of pitch and plunge, and the second is the other pair dominated by pitch and plunge, retained because the
limit-cycle amplitude depends on it (Section~\ref{sec:accuracy}).
Table~\ref{tab:modes} lists the spectrum at the construction point $\Uref = 7.0$, where $\lambda$ is an eigenvalue of $\vect{A}$, not the decay rate of
Eq.~\eqref{eq:circuit}. The flap pair oscillates at a reduced frequency of
$0.578$, nearly eight times that of the flutter mode, and the flap deflection in
the limit cycle is small. The Wagner and K\"ussner poles are real, strongly damped
and coupled to the structural motion, and each Wagner pole appears once per degree of
freedom, and structural coupling displaces one of each triple. The transducer pole
is displaced by the coupling from $-1/(\Us r) = -0.1429$ to $-0.1385$, next to the
K\"ussner pole $-\varepsilon_3$. Section~\ref{sec:accuracy} tests the omission of
the flap pair and of the electrical modes by adding each to the basis.

\begin{table}[H]
\centering
\caption{Eigenvalues of the fifteen-state Jacobian $\vect{A}(\Us)$ at $\Us = 7.0$,
plunge mounting, $\chi = 0.20$, $r = 1.0$, per unit $\tau$.}
\label{tab:modes}
\small
\begin{tabular}{llll}
\toprule
eigenvalue $\lambda$ & type & mechanism & retained \\
\midrule
$+0.02009 \pm 0.07377\mathrm{i}$ & pair & critical flutter mode, pitch-plunge & yes \\
$-0.07196 \pm 0.05636\mathrm{i}$ & pair & second aeroelastic mode, pitch-plunge & yes \\
$-0.01234 \pm 0.57760\mathrm{i}$ & pair & flap mode & no \\
$-0.13846$ & real & transducer pole (electrical state) & no \\
$-0.13930$ & real & K\"ussner gust lag, $\varepsilon_3$ & no \\
$-1.80200$ & real & K\"ussner gust lag, $\varepsilon_4$ & no \\
$-0.02529$ & real & Wagner lag, $\varepsilon_1$ branch, displaced & no \\
$-0.27142$ & real & Wagner lag, $\varepsilon_2$ branch, displaced & no \\
$-0.04550$ & real, $\times2$ & Wagner lag, $\varepsilon_1$ & no \\
$-0.30000$ & real, $\times2$ & Wagner lag, $\varepsilon_2$ & no \\
\bottomrule
\end{tabular}
\end{table}

Marching Eq.~\eqref{eq:reduced} away from the construction point needs the reduced
operator at the new velocity, and the two treatments compared throughout differ
only in how it is obtained. The established treatment expands in $\dU = \Us-\Uref$.
What it carries in $\Us$ is each retained eigenvalue, and the reduced operator is
reassembled from them,
\begin{equation}
\lambda_{i}(\Us) = \sum_{k=0}^{K}\frac{1}{k!}
\left.\frac{\mathrm{d}^{k}\lambda_{i}}{\mathrm{d}\Us{}^{k}}\right|_{\Uref}
(\dU)^{k},
\qquad
\vect{A}_{r}(\Us) = \operatorname{blkdiag}_{i}
\begin{bmatrix}
\mathrm{Re}\,\lambda_{i} & -\mathrm{Im}\,\lambda_{i}\\
\mathrm{Im}\,\lambda_{i} & \phantom{-}\mathrm{Re}\,\lambda_{i}
\end{bmatrix},
\label{eq:aexp}
\end{equation}
with $K=3$ in every result reported here, lower orders being used only where the
effect of the order is examined. The derivatives are formed at construction time
by central differences in $\Us$, each re-solving the eigenproblem, so the basis
and every reduced operator stay frozen. The alternative re-projects the exact
Jacobian onto the same frozen basis,
\begin{equation}
\vect{A}_{r}(\Us) = \vect{W}^{\mathsf T}\,\vect{A}(\Us)\,\vect{V},
\label{eq:areproj}
\end{equation}
at the cost of one Jacobian evaluation and two matrix products per velocity,
leaving the matrix-free construction of $\vect{C}_r$ untouched. The two coincide
at $\Us=\Uref$. Their structural difference, which turns out to govern the
accuracy, is that Eq.~\eqref{eq:areproj} yields a general matrix whereas
Eq.~\eqref{eq:aexp} yields a block-diagonal one at every velocity and every order
$K$. The expansion moves each retained eigenvalue along its own curve, but it
cannot couple the retained modes, because no off-block quantity is ever computed.
Reduced models are labelled by these equation numbers below.

\section{Time integration and measurement of the limit cycle}
\label{sec:march}

Every response, full-order and reduced, is obtained by the same procedure. The
system is marched from rest by Heun's explicit second-order Runge-Kutta scheme with
a fixed step of $0.05$ in $\tau$, about $1500$ steps per cycle. The initial
condition is a pitch angle of $10^{-3}$~rad, about $0.06^\circ$, with plunge, flap,
all three rates, the eight lag states and the voltage at zero, and the reduced model starts from the projection of that state onto its basis. 
No gust is applied, so the response either decays to the equilibrium or grows onto a limit cycle. Convergence
is decided in blocks of $20000$ steps, or $1000$ time units. A response is a
converged limit cycle when the largest pitch amplitude in a block differs from that
in the preceding block by less than $0.1\%$, and has decayed when the equilibrium
is linearly stable and the amplitude is falling, or when it has fallen below
$10^{-9}$. The cap is $40$ blocks, scaled down where the growth rate allows, and a
case that reaches it is reported as unconverged. Amplitudes, period and mean power
are measured on the last block over a whole number of cycles, between the first and
the last upward zero crossing of the voltage, so that no partial cycle biases the
mean.

\section{Results}
\label{sec:results}

\subsection{Verification}

Three limiting cases confirm the implementation over $5\le\Us\le10$ and
$r = 0.1$, $1$ and $10$. With the harvester disabled the coupled Jacobian and
residual reproduce the baseline fourteen-state system exactly. With the harvester
enabled but $\chi = 0$, the spectrum is the fourteen baseline eigenvalues,
recovered to within $1.8\times10^{-15}$, plus a single real eigenvalue at
$-1/(\Us r)$, recovered exactly, and the harvested power reduces exactly to
$\chi v^2/r$.

\subsection{Effect of the transducer on the flutter boundary}

\begin{figure}[H]
\centering
\includegraphics[width=\textwidth]{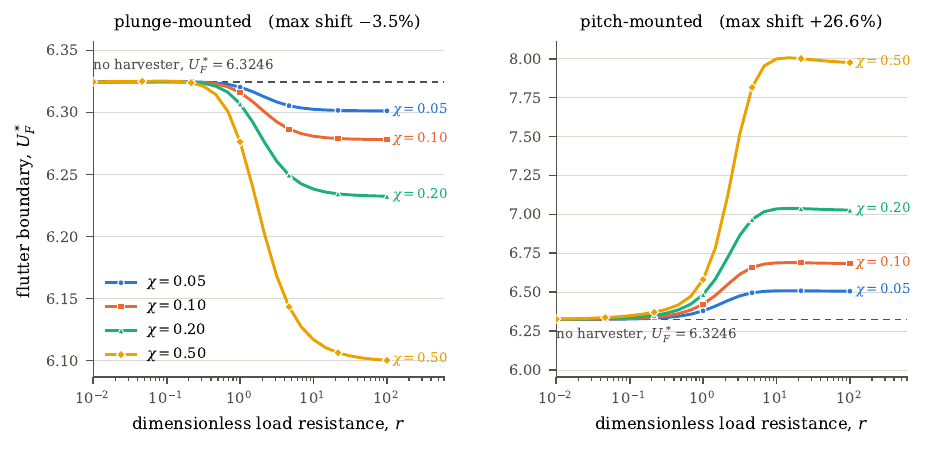}
\caption{Linear flutter boundary against load resistance, transducer on the plunge
(left) and pitch (right) degree of freedom. The dashed line is the baseline.}
\label{fig:A}
\end{figure}

The flutter boundary is the lowest reduced velocity at which the largest real part
of the linearised spectrum crosses zero. The baseline flutter boundary is at
$\Us_F = 6.3246$, and the transducer shifts it in a direction set by the degree of freedom that carries it (Table~\ref{tab:flutter},
Figure~\ref{fig:A}). A plunge-mounted transducer is destabilising, lowering the
boundary monotonically with load resistance to $6.1002$ ($-3.5\%$) at $\chi = 0.50$
in the open-circuit limit. A pitch-mounted transducer is stabilising, raising it to
$8.0058$ ($+26.6\%$) at $\chi = 0.50$ and $r = 14.7$. Both shifts vanish in the
short-circuit limit, since a transducer held at zero voltage exerts no force, and
both mountings recover the baseline as $r\to0$. They do not vanish in the
open-circuit limit. The plunge mounting approaches its largest shift there, and
the pitch mounting passes through an interior optimum at $r = 14.7$ and still
stands at $+26.1\%$ at $r = 100$. An open-circuited transducer delivers no power
but acts as an added stiffness, which is what displaces the boundary.

\begin{table}[H]
\centering
\caption{Linear flutter boundary $\Us_F$ against load resistance $r$ for the
plunge- and pitch-mounted transducer at four couplings $\chi$.}
\label{tab:flutter}
\small
\begin{tabular}{llrrrrr}
\toprule
mounting & $\chi$ & \multicolumn{5}{c}{$\Us_F$ at load resistance $r$} \\
\cmidrule(lr){3-7}
 & & $r = 0.01$ & $r = 0.1$ & $r = 1$ & $r = 10$ & $r = 100$ \\
\midrule
plunge & 0.05 & 6.3246 & 6.3247 & 6.3204 & 6.3026 & 6.3013 \\
plunge & 0.10 & 6.3246 & 6.3247 & 6.3161 & 6.2808 & 6.2781 \\
plunge & 0.20 & 6.3246 & 6.3249 & 6.3069 & 6.2382 & 6.2325 \\
plunge & 0.50 & 6.3247 & 6.3252 & 6.2763 & 6.1168 & 6.1002 \\
\addlinespace
pitch & 0.05 & 6.3249 & 6.3279 & 6.3801 & 6.5090 & 6.5067 \\
pitch & 0.10 & 6.3252 & 6.3310 & 6.4232 & 6.6886 & 6.6843 \\
pitch & 0.20 & 6.3258 & 6.3365 & 6.4850 & 7.0352 & 7.0274 \\
pitch & 0.50 & 6.3274 & 6.3489 & 6.5812 & 7.9987 & 7.9753 \\
\bottomrule
\end{tabular}
\end{table}

\subsection{Limit-cycle oscillation and harvested power}

\begin{figure}[H]
\centering
\includegraphics[width=0.94\textwidth]{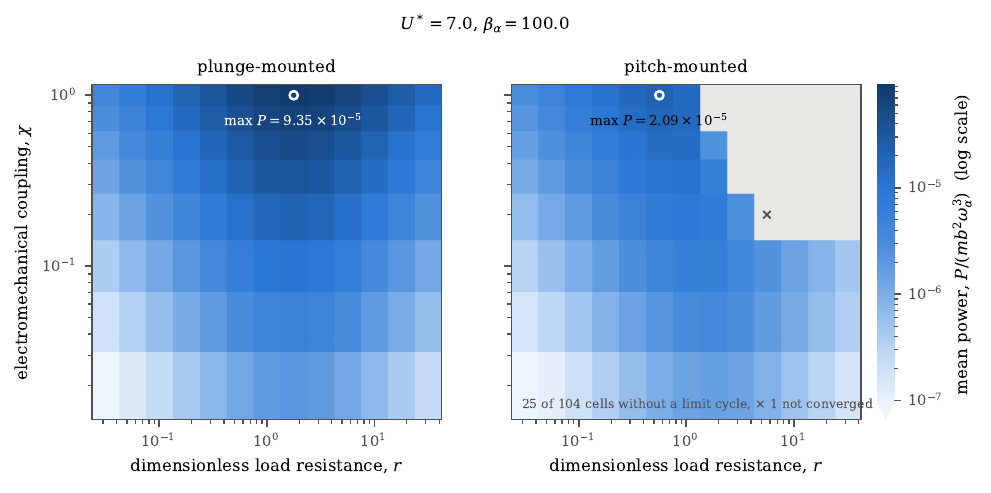}
\caption{Mean harvested power over the coupling / load-resistance plane at
$\Us = 7.0$, plunge (left) and pitch (right) mounting, on the common reference
$P/(mb^2\omega_\alpha^3)$. Grey cells have no converged limit cycle.}
\label{fig:B}
\end{figure}

Above the flutter boundary the response settles onto a limit cycle whose amplitude
grows as the square root of the excess velocity, the signature of a supercritical
Hopf bifurcation, and whose mean power grows, to leading order near the
bifurcation, as the square of that amplitude. The mean power over the coupling /
load-resistance plane at $\Us = 7.0$ (Figure~\ref{fig:B}) has a single interior
optimum in $r$ for every coupling, moving to lower $r$ as $\chi$ increases. On the
common mechanical reference the plunge mounting reaches $9.35\times10^{-5}$ at
$\chi = 1.00$, $r = 1.78$ and the pitch mounting $2.09\times10^{-5}$ at
$\chi = 1.00$, $r = 0.56$, a factor of $4.5$ in favour of plunge. Of the $104$
pitch-mounted cells, $78$ reach a converged limit cycle, $25$ are linearly stable
and support none, and one, at $\chi = 0.20$, $r = 5.623$, is linearly unstable with
a growth rate of $7.9\times10^{-5}$ but had not converged within the marching
budget. All $26$ are masked rather than reported as low power.

\subsection{Accuracy of the reduced model}
\label{sec:accuracy}

\begin{figure}[H]
\centering
\includegraphics[width=\textwidth]{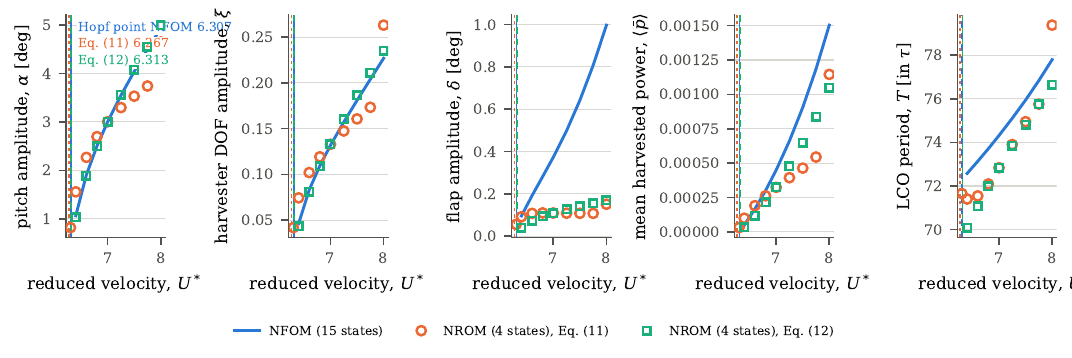}
\caption{Four-state NROM against the fifteen-state NFOM over reduced velocity,
Eq.~\eqref{eq:aexp} (circles) and Eq.~\eqref{eq:areproj} (squares), basis at
$\Us = 7.0$. Dashed lines mark the Hopf points.}
\label{fig:C}
\end{figure}

\begin{figure}[H]
\centering
\includegraphics[width=0.9\textwidth]{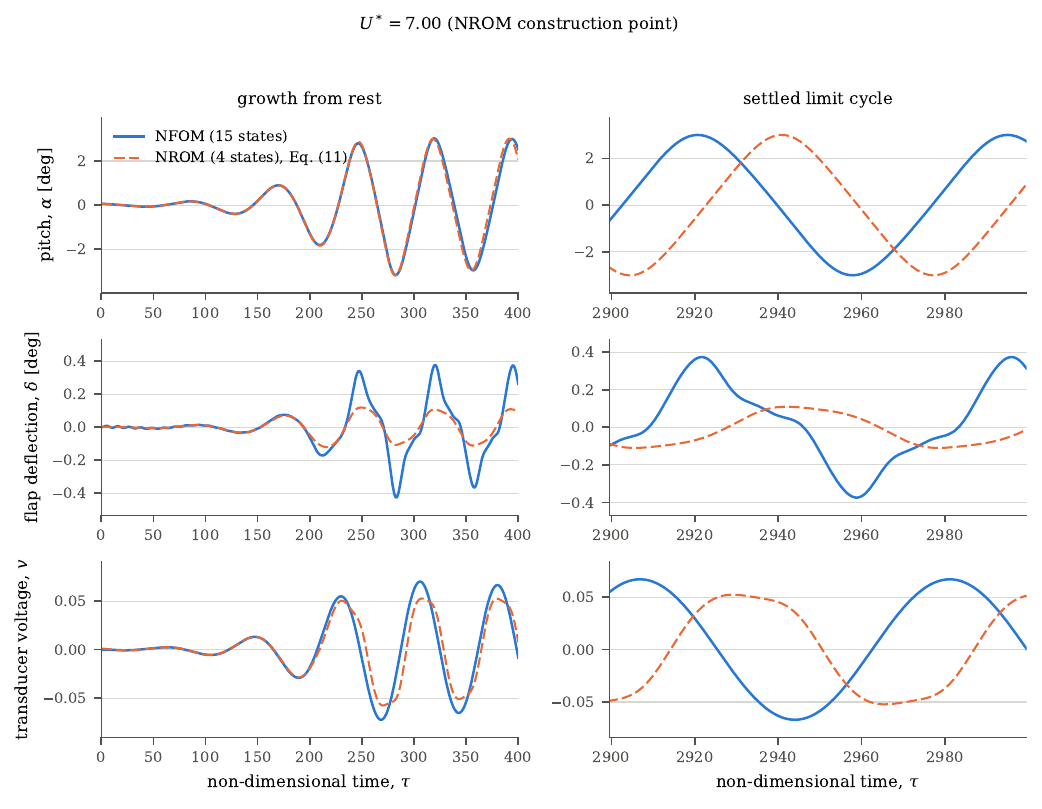}
\caption{Pitch, flap deflection and transducer voltage at $\Us = 7.0$ during growth
from rest (left) and on the settled limit cycle (right).}
\label{fig:D}
\end{figure}

Accuracy is assessed separately at the construction point, where only the modal
truncation acts, and away from it, where the velocity treatment also enters. All
results use the plunge mounting at $\chi = 0.20$, $r = 1.0$, $\beta_\alpha = 100$,
with the basis built at $\Uref = 7.0$. Figure~\ref{fig:C} and
Table~\ref{tab:romfom} compare the limit-cycle amplitudes, the mean power and the
period of the two reduced operators with those of the NFOM across the range, and
Figure~\ref{fig:D} gives the time histories at the construction point.

At the construction point the four-state NROM reproduces the NFOM pitch amplitude
to $+0.1\%$ and the plunge amplitude to $+0.8\%$, whereas retaining the critical pair alone
gives $+11.5\%$ and $+72.2\%$, so the second retained pair is necessary. The NROM
period is about $2\%$ short, which accumulates into a phase drift of about $0.3$ of
a cycle by the settled window of Figure~\ref{fig:D}. The transducer voltage is
$21.9\%$ low. The transducer pole and the K\"ussner pole lie within
$8.4\times10^{-4}$ of each other, so their eigenvectors mix and must be retained
together, and adding both improves the voltage only to $-16.0\%$ while leaving
pitch and plunge essentially unchanged. Retaining the full spectrum recovers every
quantity exactly as anticipated.

The settled voltage is very nearly a pure sinusoid, with a crest factor of $1.412$
against $1.414$ for a sine and a third harmonic of $0.2\%$ of the fundamental
against $3.8\%$ for pitch, so the deficit is not due to harmonic content outside
the retained modes. It lies in the proportion of voltage to displacement carried by
the basis. Projecting the converged NFOM cycle onto the four-state subspace keeps
pitch and plunge to within $3.0\%$ and $6.4\%$ but itself loses $15.4\%$ of the
voltage, and the reduced dynamics then lose a further $7.7\%$. On any eigenvector
of the equilibrium Jacobian, Eq.~\eqref{eq:circuit} fixes the voltage-to-plunge
ratio at the circuit gain $|s|/|s+\lambda|$ for motion at that eigenvalue, with
$\lambda = 1/(\Us r) = 0.1429$. For the critical pair,
$s = 0.0201 + 0.0738\mathrm{i}$, the gain is $0.427$, and the eigenvector carries
exactly that ratio. The limit cycle runs with zero growth at the reduced frequency
$\omega_0 = 0.0845$, where the gain is $\omega_0/|\mathrm{i}\omega_0+\lambda| = 0.509$,
and the NFOM cycle carries $0.508$. No combination of structural eigenvectors can
alter the proportion, so the projected cycle carries only $0.427/0.509$ of the
voltage, a predicted shortfall of $16\%$ against the $15.4\%$ measured. The NROM's
own cycle within that subspace is smaller again, $0.0523$ against the $0.0567$
that the projected NFOM cycle carries, because the neglected modes no longer feed
the retained ones. Only a state for the voltage itself, supplied by the electrical
modes, can alter the proportion, and those modes together carry only $0.026$ of the
voltage on the NFOM cycle, which is why their inclusion helps only in part.

Away from the construction point the error is dominated by the velocity treatment.
Carrying the reduced operator by Eq.~\eqref{eq:aexp} displaces the reduced Hopf
point to $6.2673$ against the full-order $6.3069$, an error of $-0.63\%$, and gives
pitch errors from $+46.4\%$ at $\Us = 6.40$, through zero at the construction
point, to $-16.9\%$ at $\Us = 7.75$, and at $\Us = 6.30$, below the full-order Hopf
point, it converges to a spurious limit cycle. Re-projecting onto the same frozen
basis, Eq.~\eqref{eq:areproj}, reduces the pitch errors to between $-3.4\%$ and
$+1.3\%$ across $6.40\le\Us\le8.00$ without enlarging the basis, and moves the Hopf
point to $6.3130$ ($+0.10\%$). Under Eq.~\eqref{eq:areproj} the voltage and power
errors are nearly flat in flow speed, a property of the basis, whereas under Eq.~\eqref{eq:aexp} the velocity treatment adds its own error. Both treatments
underpredict the period at every velocity, by $1.4$ to $3.4\%$ under
Eq.~\eqref{eq:areproj} and by $1.3$ to $2.2\%$ under Eq.~\eqref{eq:aexp}, except at
$\Us = 8.00$ under Eq.~\eqref{eq:aexp}. There the errors change sign and the period
jumps to $79.38$, with alternating peaks and cycle lengths, so the solution is of
period two rather than the simple limit cycle the convergence test reports.
Reporting truncation and velocity extrapolation together, as one error attributed
to the reduction, therefore overstates the accuracy lost by reducing the model by
an order of magnitude.

\begin{table}[H]
\centering
\caption{NROM against NFOM over reduced velocity ($\chi = 0.20$, $r = 1.0$, basis
at $\Uref = 7.0$). Column~2 is the NFOM pitch amplitude and the others are relative
errors in $\alpha$, $\xi$, $\delta$, $v$ and $\langle\bar p\rangle$.}
\label{tab:romfom}
\small
\setlength{\tabcolsep}{3.2pt}
\begin{tabular}{lr rrrrr rrrrr}
\toprule
& & \multicolumn{5}{c}{NROM, Eq.~\eqref{eq:aexp} [\%]}
  & \multicolumn{5}{c}{NROM, Eq.~\eqref{eq:areproj} [\%]} \\
\cmidrule(lr){3-7}\cmidrule(lr){8-12}
$\Us$ & $\alpha_{\mathrm{NFOM}}$ [deg]
 & $\alpha$ & $\xi$ & $\delta$ & $v$ & $\langle\bar p\rangle$
 & $\alpha$ & $\xi$ & $\delta$ & $v$ & $\langle\bar p\rangle$ \\
\midrule
6.20 & \multicolumn{11}{l}{\emph{NFOM linearly stable and both reduced operators decay, so no limit cycle}} \\
6.30 & \multicolumn{11}{l}{\emph{NFOM and Eq.~\eqref{eq:areproj} decay, while Eq.~\eqref{eq:aexp} converges to a spurious limit cycle}} \\
6.40 & 1.06 & $+46.39$ & $+61.13$ & $-4.38$ & $+44.07$ & $+100.17$ & $-3.40$ & $-6.51$ & $-57.63$ & $-19.90$ & $-33.92$ \\
6.60 & 1.91 & $+18.84$ & $+22.75$ & $-43.69$ & $+7.16$ & $+13.98$ & $-0.91$ & $-2.54$ & $-62.07$ & $-19.88$ & $-29.64$ \\
6.80 & 2.50 & $+7.72$ & $+8.92$ & $-60.98$ & $-9.83$ & $-12.63$ & $-0.28$ & $-0.64$ & $-66.89$ & $-21.10$ & $-28.38$ \\
7.00 & 3.00 & $+0.10$ & $+0.77$ & $-70.65$ & $-21.93$ & $-27.87$ & $+0.10$ & $+0.77$ & $-70.65$ & $-21.93$ & $-27.87$ \\
7.25 & 3.55 & $-7.09$ & $-6.15$ & $-78.17$ & $-31.32$ & $-40.39$ & $+0.47$ & $+2.08$ & $-74.58$ & $-22.39$ & $-27.85$ \\
7.50 & 4.04 & $-12.62$ & $-11.19$ & $-83.12$ & $-31.63$ & $-48.72$ & $+0.79$ & $+2.97$ & $-77.89$ & $-22.59$ & $-28.32$ \\
7.75 & 4.50 & $-16.85$ & $-15.02$ & $-86.61$ & $-30.71$ & $-54.01$ & $+1.08$ & $+3.48$ & $-80.67$ & $-22.79$ & $-29.15$ \\
8.00 & 4.93 & $-5.74$ & $+15.98$ & $-84.84$ & $+31.67$ & $-23.83$ & $+1.34$ & $+3.64$ & $-83.01$ & $-23.10$ & $-30.29$ \\
\bottomrule
\end{tabular}
\end{table}

The four-state basis also misrepresents the flap deflection. Table~\ref{tab:basis}
isolates the effect of the basis by comparing reduced models at the construction
point, where the two velocity treatments coincide, each marched from the same
initial condition as the NFOM and measured in the same way. The flap amplitude is
$0.37^\circ$ against $3.00^\circ$ of pitch, and the four-state basis gives $70.6\%$
too little of it. Under Eq.~\eqref{eq:areproj} the error runs from
$-58\%$ at $\Us = 6.40$ to $-83\%$ at $\Us = 8.00$ while the NFOM flap amplitude
grows from $0.10^\circ$ to $1.0^\circ$, so the shortfall is a property of the basis
rather than of the extrapolation. The flap mode, at
$\lambda = -0.0123 \pm 0.5776\mathrm{i}$, has a frequency $7.8$ times that of the flutter mode,
but in the limit cycle the flap follows the pitch and plunge motion at the flutter
frequency, driven through the hinge moment, and the retained eigenvectors carry only
a small flap component. The flap motion also carries a third harmonic of $29\%$ of
its fundamental, from the cubic pitch spring acting through the aerodynamic
coupling. Unlike the voltage deficit, this is an ordinary truncation error. Retaining
the flap pair as well, resulting in six states, brings the flap to $-3.9\%$ and leaves pitch,
plunge and voltage at $+0.1\%$, $+1.3\%$ and $-21.5\%$. Retaining it in place of
the second pair leaves pitch and plunge at $+11.5\%$ and $+73.2\%$, since that pair
determines them, and overpredicts the flap by $47.5\%$. Adding the electrical modes
on top, eight states, gives $-3.3\%$ in the flap and $-15.5\%$ in the voltage, so
the two deficits are independent and only the electrical one persists as the basis
is enlarged. In Figure~\ref{fig:D} the NROM reproduces the shape of the flap motion
at less than a third of its amplitude.

\begin{table}[H]
\centering
\caption{Relative error of the NROM, $100\,(\mathrm{NROM}-\mathrm{NFOM})/\mathrm{NFOM}$,
in the limit-cycle amplitudes at $\Uref = 7.0$ for each retained basis, with $q$ the number of real reduced coordinates.}
\label{tab:basis}
\small
\begin{tabular}{p{6.4cm}rrrrr}
\toprule
eigenvectors retained in the basis & $q$ & $\alpha$ [\%] & $\xi$ [\%] & $\delta$ [\%] & $v$ [\%] \\
\midrule
critical pair only & 2 & $+11.49$ & $+72.22$ & $-41.68$ & $+27.25$ \\
critical and second aeroelastic pairs (used throughout) & 4 & $+0.10$ & $+0.77$ & $-70.65$ & $-21.93$ \\
critical and flap pairs & 4 & $+11.50$ & $+73.23$ & $+47.50$ & $+28.02$ \\
critical, second aeroelastic and flap pairs & 6 & $+0.14$ & $+1.29$ & $-3.92$ & $-21.50$ \\
critical pair and the two electrical modes & 4 & $+11.79$ & $+74.68$ & $-40.31$ & $+43.35$ \\
critical and second aeroelastic pairs and the two electrical modes & 6 & $+0.23$ & $+2.11$ & $-70.17$ & $-15.99$ \\
critical, second aeroelastic and flap pairs and the two electrical modes & 8 & $+0.27$ & $+2.63$ & $-3.27$ & $-15.54$ \\
full spectrum & 14 & $0.00$ & $0.00$ & $0.00$ & $0.00$ \\
\bottomrule
\end{tabular}
\end{table}

\subsection{Cost of the reduction}
\label{sec:cost}

At this model scale and aerodynamic fidelity, the reduction yields no significant time integration advantage. 
Over $1000$ units of $\tau$ the ratio of march times depends almost entirely on how carefully each
residual is written. Against the full-order residual as first written the
four-state march appears $1.5$ times faster, but giving the full-order residual the
same treatment, with its arithmetic verified unchanged, moves the ratio to between
$1.0$ and $1.4$. Neither march is limited by arithmetic. One full-order residual
performs about $510$ floating-point operations, some $0.4\%$ of its measured cost,
and the residual cost is flat in the state count from eight to one hundred states,
so reduction begins to pay only at order one hundred states. The projected cubic
operator is moreover dense, with $(2m)^4$ entries, and at ten reduced states
already costs more than the fifteen-state residual. Over the sweep the reduced
marches also need $1.5$ times (Eq.~\eqref{eq:aexp}) and $2.0$ times
(Eq.~\eqref{eq:areproj}) as much simulated time as the full-order ones. Under
Eq.~\eqref{eq:areproj} all of that excess lies in the two linearly stable cases
just below the Hopf point, where the full-order driver stops on the spectrum while
the reduced march waits for the transient to decay, whereas under Eq.~\eqref{eq:aexp} its
spurious cycle at $\Us = 6.30$ and slower convergence at $7.75$ and $8.00$ add to
it. Where all three models reach a limit cycle promptly, $6.40\le\Us\le7.50$, they
take the same simulated time.

\subsection{The linear measure does not reveal the error}
\label{sec:linear}

\begin{figure}[H]
\centering
\includegraphics[width=\textwidth]{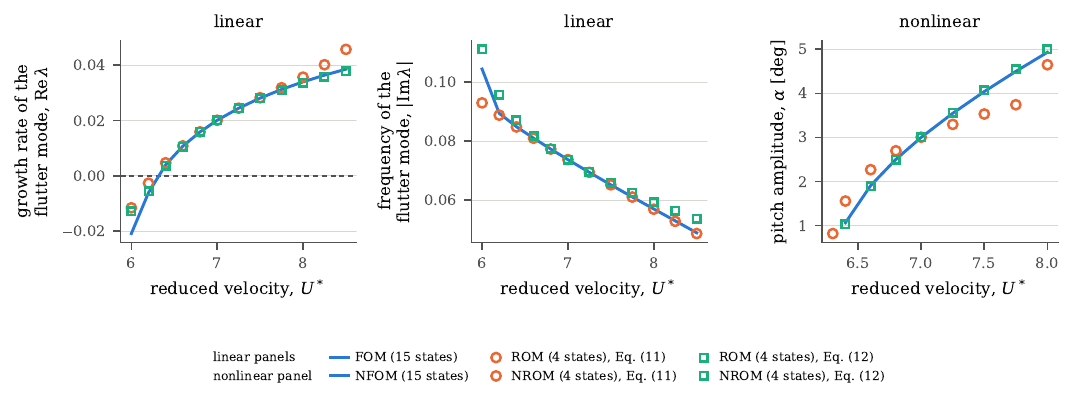}
\caption{Growth rate and frequency of the flutter branch, ROM against FOM (left,
centre), and limit-cycle pitch amplitude, NROM against NFOM (right).}
\label{fig:E}
\end{figure}

Table~\ref{tab:linear} compares the two velocity treatments on the linearised
system, following the flutter branch by continuation from the construction point,
since below the boundary the least-stable FOM eigenvalue belongs to the flap mode.
Over $6.0\le\Us\le8.5$ the largest growth-rate error is $9.3\times10^{-3}$ for
Eq.~\eqref{eq:aexp} and $8.0\times10^{-3}$ for Eq.~\eqref{eq:areproj}, essentially
the same, yet the nonlinear amplitude errors of Table~\ref{tab:romfom} differ by
more than an order of magnitude (Figure~\ref{fig:E}). Near a Hopf bifurcation the
limit-cycle amplitude scales as the square root of the excess growth rate, so a
small linear error is strongly amplified in the nonlinear response, and a reduced
model validated on its linear spectrum alone can look adequate while ill-predicting the
limit cycle dynamic region.

\begin{table}[H]
\centering
\caption{Growth rate of the flutter branch, $\mathrm{Re}\,\lambda$ per unit $\tau$,
for the FOM and the two reduced operators (basis at $\Uref = 7.0$).}
\label{tab:linear}
\small
\begin{tabular}{lrrr}
\toprule
$\Us$ & FOM & ROM, Eq.~\eqref{eq:aexp} & ROM, Eq.~\eqref{eq:areproj} \\
\midrule
6.00 & $-0.020817$ & $-0.011514$ & $-0.012782$ \\
6.20 & $-0.006049$ & $-0.002642$ & $-0.005597$ \\
6.40 & $+0.004043$ & $+0.004743$ & $+0.003746$ \\
6.60 & $+0.010748$ & $+0.010853$ & $+0.010617$ \\
6.80 & $+0.015893$ & $+0.015898$ & $+0.015864$ \\
7.00 & $+0.020091$ & $+0.020091$ & $+0.020091$ \\
7.25 & $+0.024445$ & $+0.024455$ & $+0.024412$ \\
7.50 & $+0.028094$ & $+0.028230$ & $+0.027978$ \\
7.75 & $+0.031225$ & $+0.031827$ & $+0.030993$ \\
8.00 & $+0.033958$ & $+0.035662$ & $+0.033587$ \\
8.25 & $+0.036376$ & $+0.040145$ & $+0.035848$ \\
8.50 & $+0.038537$ & $+0.045690$ & $+0.037840$ \\
\midrule
\multicolumn{2}{l}{max.\ absolute deviation from FOM}
 & $9.3\times 10^{-3}$ & $8.0\times 10^{-3}$ \\
\bottomrule
\end{tabular}
\end{table}

\subsection{Why the velocity expansion fails}
\label{sec:coupling}

Because Eq.~\eqref{eq:aexp} reassembles the reduced operator from the retained
eigenvalues alone, its inter-mode entries are identically zero at every velocity
and every order. Taking Eq.~\eqref{eq:areproj}, which is exact on this basis, as the
reference, the difference is split into the retained-mode blocks, whose error is
the truncation error of the expansion and the only part its order affects, and the
inter-mode blocks, which Eq.~\eqref{eq:aexp} sets to zero (Table~\ref{tab:coupling}), and both vanish to machine precision at
$\Us = 7.0$, where the two operators coincide.
The discarded coupling is not a small correction. It reaches $8.6\times10^{-2}$ at
$\Us = 8.5$, comparable to the retained eigenvalues themselves, $0.020$ and $0.074$
in real and imaginary part at the construction point. It is also consistently about
twice the truncation error in the retained-mode blocks, with a median ratio of
$1.9$, so the larger part of the error in Eq.~\eqref{eq:aexp} is the part no
additional expansion term can reach.

\begin{table}[H]
\centering
\caption{Largest absolute entry, per unit $\tau$, of the error of
Eq.~\eqref{eq:aexp} ($K = 3$) relative to Eq.~\eqref{eq:areproj}, in the
retained-mode and inter-mode blocks.}
\label{tab:coupling}
\small
\begin{tabular}{lrrr}
\toprule
$\Us$ & retained-mode blocks & inter-mode blocks & ratio \\
\midrule
6.20 & $3.81\times10^{-2}$ & $7.35\times10^{-2}$ & 1.9 \\
6.40 & $2.67\times10^{-2}$ & $5.25\times10^{-2}$ & 2.0 \\
6.60 & $1.67\times10^{-2}$ & $3.34\times10^{-2}$ & 2.0 \\
6.80 & $7.78\times10^{-3}$ & $1.60\times10^{-2}$ & 2.1 \\
7.00 & $1.60\times10^{-16}$ & $4.92\times10^{-16}$ & --- \\
7.25 & $8.93\times10^{-3}$ & $1.81\times10^{-2}$ & 2.0 \\
7.50 & $1.77\times10^{-2}$ & $3.45\times10^{-2}$ & 1.9 \\
7.75 & $2.66\times10^{-2}$ & $4.92\times10^{-2}$ & 1.9 \\
8.00 & $3.56\times10^{-2}$ & $6.27\times10^{-2}$ & 1.8 \\
8.25 & $4.50\times10^{-2}$ & $7.49\times10^{-2}$ & 1.7 \\
8.50 & $5.49\times10^{-2}$ & $8.60\times10^{-2}$ & 1.6 \\
\bottomrule
\end{tabular}
\end{table}

Raising the expansion order does not recover the coupling. From second to third order the
inter-mode coupling stays at exactly zero and the growth rate is carried across the
full-order value rather than onto it. At $\Us = 8.5$, against a full-order growth
rate of $0.038537$, second order gives $0.030829$ and third order $0.045690$, an overshoot
of $0.0072$ against an undershoot of $0.0077$, so the extra term reverses the sign
of the error and takes only $7\%$ off its magnitude. In the retained-mode blocks
the order dependence is sided. Going from first to second order lowers the error at
all ten velocities tested, whereas the third-order term lowers it at the four below
the construction point and raises it at the six above, to $5.49\times10^{-2}$ against
the second-order $4.99\times10^{-2}$ at $\Us = 8.5$, and the growth rate, one scalar
summary of the block, can improve while its entries grow worse. The series thus
stops converging on the side that matters, since a limit cycle exists only above
the Hopf point, and on the other side converges to an operator with no inter-mode
coupling. The two measures of Section~\ref{sec:linear} therefore differ for a structural reason. The linear
growth rate depends on the diagonal blocks, which the eigenvalue continuation
represents well, whereas the limit-cycle amplitude depends on how the retained modes
exchange energy, which lives in the off-blocks that Eq.~\eqref{eq:aexp} sets to
zero. The remedy for this error is therefore neither a larger basis nor a higher
order, but the single triple product of Eq.~\eqref{eq:areproj}. The voltage deficit is a property of the basis and is
not affected by this remedy.

\section{Discussion}
\label{sec:discussion}

The mounting degree of freedom is a primary design variable. The plunge mounting delivers four
and a half times the power of the pitch mounting on a common reference, and the two
shift the flutter boundary in opposite directions. A stabilising mounting raises the
speed at which harvesting begins, which suits structural integrity but not energy
capture, while the destabilising mounting that harvests better also erodes the
flutter margin. The sign of that trade is fixed by where the transducer is placed,
not by how strongly it is coupled.

Reported as one error over a range of flow speeds, the NROM pitch error here is
$+46\%$ to $-17\%$, which would suggest a basis far too small. If decomposed, it is
sufficient, since at the construction point the same four-state basis is correct to
$0.1\%$ in pitch and $0.8\%$ in plunge. The large errors arise away from it, from an
expansion whose inter-mode coupling is zero at every order while the exact operator
carries a coupling comparable to the eigenvalues themselves. The spectrum is set by
the diagonal blocks, whereas the limit cycle is set by the coupling. 
Re-projection repairs it at the cost of one Jacobian evaluation and two matrix products per velocity and removes the spurious limit cycle
below the Hopf point. Truncation error and velocity-treatment error should
therefore be reported separately, because they are controlled by different choices.

With the velocity treatment corrected, the residual error lies in the harvested
quantity. The voltage stays $20$ to $23\%$ low and the mean power $28$ to $34\%$
low, nearly independently of flow speed, while pitch and plunge are right to a few
per cent. Each equilibrium eigenvector carries the voltage in the proportion the
circuit gain takes at its own eigenvalue, whereas the limit cycle runs at a higher
frequency with zero growth, where the gain is $19\%$ larger, and no combination of
structural eigenvectors can change that proportion. The flap deflection, missed by
$58$ to $83\%$, is by contrast an ordinary truncation error that the flap pair
removes at six states. A harvester ROM validated on its structural response would
therefore appear an order of magnitude more accurate than it is for the quantity
being designed for.

The aerodynamics assume attached, incompressible, two-dimensional flow, so the
limit cycle is held below five degrees and no claim is made about stall-dominated
LCOs. The transduction is linear with a purely resistive load, with no rectifier,
storage or impedance matching, so the powers are upper bounds on what a real circuit
would deliver. Results are non-dimensional, which makes the comparison between
mountings exact but not directly comparable with the dimensional figures
of~\citep{erturk2010,dunnmon2011}, and a dimensional study against the experimental
configuration of~\citep{fichera2014} is the natural next step. The cost conclusion
is specific to an interpreted implementation at fifteen states.

\section{Conclusions}
\label{sec:conclusions}

A piezoelectric transducer was embedded in a three-degree-of-freedom pitch-plunge
aerofoil with a finite-mass flap and unsteady strip-theory aerodynamics, and a
four-state NROM of the fifteen-state system was built by matrix-free projection of
the Taylor-expanded residual.
\begin{enumerate}[leftmargin=*,itemsep=2pt]
\item Aeroelastic limit cycles are a usable energy source, and the mounting degree
of freedom is a first-order design variable. It sets the sign of the
flutter-boundary shift, by up to $-3.5\%$ for plunge and up to $+26.6\%$ for
pitch, and changes the harvested power by a factor of $4.5$ in favour of plunge.
\item The accuracy of the reduced model is dominated by how the reduced operator is
made to depend on flow speed, not by the size of the basis. Exact re-projection
onto the same frozen basis, rather than expansion of its eigenvalues, reduces the
peak pitch-amplitude error from $46\%$ to $3.4\%$ at negligible cost.
\item That failure is structural. An expansion carried through the retained
eigenvalues has identically zero inter-mode coupling at every order, so it
reproduces the linear spectrum but not the limit cycle, and no number of expansion
terms repairs it.
\item The error that remains lies in the harvested quantity. With pitch and plunge
correct to a few per cent and the flap recoverable to within $4\%$ by retaining its
own pair, the voltage is still $20$ to $23\%$ low, because each equilibrium
eigenvector fixes the voltage-to-displacement ratio at the circuit gain for its own
eigenvalue, below the gain on the periodic cycle. Harvester ROM accuracy must
therefore be reported on the harvested quantity itself.
\item At fifteen states the reduction buys no wall-clock time, because both marches
are limited by array-operation dispatch rather than arithmetic, and the cost argument
belongs to full-order models an order of magnitude larger. 
However, future studies will focus on much larger scale systems including higher fidelity aerodynamics and investigate the benefit of the model order reduction there in the limit cycle dynamic region.
\end{enumerate}

\section*{Data availability}

All data generated in this study, the scripts that produce every figure and table,
and the source code of the aeroelastic solver are available from the corresponding
author upon reasonable request.

\section*{Funding and competing interests}

The authors received no financial support for this work and declare no competing
interests.

\section*{Declaration of generative AI use}

Claude (Anthropic) was used to assist in drafting and editing the text, refining
its language and proofreading it, and in checking the analysis scripts and the
consistency of the text with the computed data. The authors reviewed all content
and take full responsibility for it.


\end{document}